\documentclass{article}
\usepackage{spconf,amsmath,graphicx,multirow}
\usepackage{url}

\title{Lossless Intra Coding in HEVC with Integer-to-Integer DST}
\name{Fatih Kamisli}
\address{Middle East Technical University\\
Ankara, Turkey}
\begin{document}
%
\maketitle
\begin{abstract}
It is desirable to support efficient lossless coding within video coding standards, which are primarily designed for lossy coding, with as little modification as possible. A simple approach is to skip transform and quantization, and directly entropy code the prediction residual, but this is inefficient for compression. A more efficient and popular approach is to process the residual block with DPCM prior to entropy coding. This paper explores an alternative approach based on processing the residual block with integer-to-integer (i2i) transforms. I2i transforms map integers to integers, however, unlike the integer transforms used in HEVC for lossy coding, they do not increase the dynamic range at the output and can be used in lossless coding. We use both an i2i DCT from the literature and a novel i2i approximation of the DST. Experiments with the HEVC reference software show competitive results.
\end{abstract}

\section{Introduction}
\label{sec:intro}

The\footnote{This research was supported by Grant 113E516 of T\"{u}bitak.} emerging high efficiency video coding (HEVC) standard \cite{HEVC} or the widely deployed H.264/AVC standard  \cite{Luthra264} support both lossy and lossless compression. Lossy compression in these standards is achieved with a block-based approach. First, a block of pixels are predicted using previously coded pixels. 
Next, the prediction error block is computed and transformed with a block based transform, and the transform coefficients are quantized and entropy coded. 

It is desirable to support efficient lossless compression using the lossy coding architecture with as little modification as possible so that encoders/decoders can also support lossless compression without any significant complexity increase. The simplest approach is to just skip the transform and quantization steps, and directly entropy code the block of prediction errors. This approach is indeed used in HEVC version 1 \cite{HEVC}. While this is a simple and low-complexity approach, it is inefficient 
and a large number of approaches have been proposed. 
A popular method is to retain the standard block-based prediction and process the residual block with differential pulse code modulation (DPCM) prior to entropy coding \cite{sulivanDPCM,cross}. 


This paper explores an alternative approach for lossless intra coding based on integer-to-integer (i2i) transforms. Integer-to-integer transforms map integers to integers. However, unlike the integer transforms used in HEVC for lossy coding, they do not increase the dynamic range at the output and can therefore be easily employed in lossless coding. While there are many papers that employ i2i approximations of the discrete cosine transform (DCT) in lossless image compression \cite{binDCT}, we could not come across a work which explores i2i transforms for lossless compression of prediction residuals in video coding standards H.264/AVC or HEVC. This paper uses an i2i DCT from the literature in lossless intra coding and presents improved coding results over the standard. In addition, this paper also derives a novel i2i approximation of the odd type-3 discrete sine transform (DST), which is the first in the literature to the best of our knowledge, and this i2i transform improves coding gains further.

The remainder of the paper is organized as follows. In Section \ref{sec:pr}, a brief overview of related previous research on lossless video compression is provided. Section \ref{sec:i2i} discusses i2i DCTs and also obtains a novel i2i approximation of the DST. Section \ref{sec:exp} presents experimental results by using these i2i transforms within HEVC for lossless intra coding. Finally, Section \ref{sec:con} concludes the paper.

\section{Previous Research}
\label{sec:pr}



\subsection{Methods based on residual DPCM}
\label{ssec:rdpcm}
Methods based on residual differential pulse code modulation (RDPCM) first perform the default block-based prediction and then process the prediction error block further with a DPCM method, i.e. a pixel-by-pixel prediction method. There are many methods proposed in the literature based on residual DPCM \cite{sulivanDPCM,cross}.

One of the earliest of such methods was proposed in \cite{sulivanDPCM} for lossless intra coding in H.264/AVC. Here, first the block-based spatial prediction is performed, and then a simple pixel-by-pixel differencing operation is applied on the residual pixels in only horizontal and vertical intra prediction modes. In horizontal mode, from each residual pixel, its left neighbor is subtracted and the result is the RDPCM pixel of the block. Similar differencing is performed along the vertical direction in the vertical intra mode. Note that the residuals of other angular modes are not processed in \cite{sulivanDPCM}. 

The same RDPCM method as in \cite{sulivanDPCM} is now included in HEVC version 2 
\cite{HEVCv2} for intra and inter coding. 
In intra coding, RDPCM is applied only when prediction mode is either horizontal or vertical.

\subsection{Methods based on modified entropy coding}
\label{ssec:entropy}
In lossy coding, transform coefficients of prediction residuals are entropy coded, while in lossless coding, the prediction residuals are entropy coded. Considering the difference of the statistics of quantized transform coefficients and prediction residuals, several modifications in entropy coding were proposed for lossless coding. 
The HEVC version 2 includes reversing the scan order of coefficients, using a dedicated context model for the significance map and other tools \cite{HEVCv2,N0044}.

\section{Integer-to-integer (i2i) transforms}
\label{sec:i2i}

Integer-to-integer (i2i) transforms map integer inputs to integer outputs and are invertible. However, unlike the integer transforms in HEVC \cite{HEVCtr}, which also map integers to integers, they do not increase the dynamic range at the output. Therefore they can be easily used in lossless compression. 

One possible method to obtain i2i transforms is to decompose a transform into a cascade of plane rotations, and then approximate each rotation with a lifting structure, which can easily map integers to integers.

\subsection{Plane rotations and the lifting scheme}
\label{ssec:rotlif}
\begin{figure}[tb]
\begin{center}
\begin{minipage}{0.29\linewidth}
\centering
\includegraphics[trim=53 680 450 55,clip,width=\linewidth]{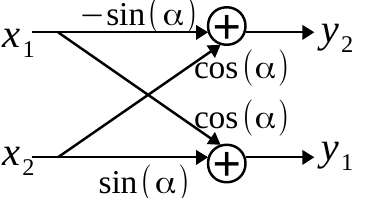}
\centerline{{\small (a) Plane rotation}}
\end{minipage}
\begin{minipage}{0.34\linewidth}
\centering
\includegraphics[trim=53 680 440 55,clip,width=\linewidth]{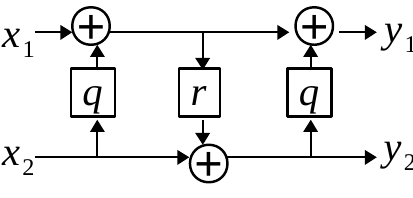}
\centerline{{\small (b) }}
\end{minipage}
\begin{minipage}{0.35\linewidth}
\centering
\includegraphics[trim=53 680 440 55,clip,width=\linewidth]{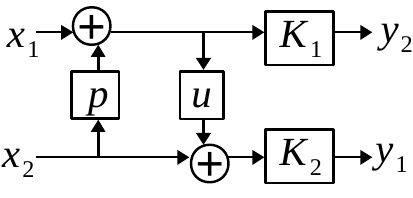}
\centerline{{\small (c) }}
\end{minipage}
\end{center}
\caption{(a) Plane rotation and its decomposition into (b) three lifting steps (c) two lifting steps and two scaling factors.}
\label{fig:rotlif32}
\end{figure}

A plane rotation can be represented with the 2x2 matrix below, and it can be decomposed into three lifting steps or two lifting steps and two scaling factors \cite{binDCT} as shown below in Equation (\ref{eq:rot2}) and in Figure \ref{fig:rotlif32}.
\begin{align}
\label{eq:rot2}
&\left[   \begin{array}{ c c }
     /cos(\alpha) &  /sin(\alpha) \\
    -/sin(\alpha) &  /cos(\alpha) 
  \end{array} \right]
= \\ \nonumber
&\left[   \begin{array}{ c c }
     1 & q \\
     0 & 1 
  \end{array} \right]
\left[   \begin{array}{ c c }
     1 & 0 \\
     r & 1 
  \end{array} \right]  
\left[   \begin{array}{ c c }
     1 & q \\
     0 & 1 
  \end{array} \right] 
= \\ \nonumber
&\left[   \begin{array}{ c c }
     0 & 1 \\
     1 & 0 
  \end{array} \right]
\left[   \begin{array}{ c c }
     K_1 & 0 \\
     0 & K_2 
  \end{array} \right]
\left[   \begin{array}{ c c }
     1 & 0 \\
     u & 1 
  \end{array} \right]  
\left[   \begin{array}{ c c }
     1 & p \\
     0 & 1 
  \end{array} \right].
\end{align}

In these equations, the lifting factors are  related to the plane rotation parameters as follows : 
\begin{itemize}
  \item $q=\frac{1-/cos(\alpha)}{/sin(\alpha)}$ and $r=-/sin(\alpha)$
  \item $p=\frac{-1}{/tan(\alpha)}$ , $u=/sin(\alpha)/cos(\alpha)$ , $K_1=-/sin(\alpha)$ and $K_2=\frac{1}{/sin(\alpha)}$.
\end{itemize}

Each lifting step can be inverted with another lifting step because 
\begin{equation}
\left[   \begin{array}{ c c }
     1 & p \\
     0 & 1 
  \end{array} \right]^{-1}
=
\left[   \begin{array}{ c c }
     1 & -p \\
     0 & 1 
  \end{array} \right]  
,
\left[   \begin{array}{ c c }
     1 & 0 \\
     u & 1 
  \end{array} \right]^{-1}
=
\left[   \begin{array}{ c c }
     1 & 0 \\
     -u & 1 
  \end{array} \right].   
\label{eq:lif}  
\end{equation}
In other words, each lifting step is inverted by subtracting out what was added in the forward lifting step. Notice that each lifting step is still invertible even if the multiplication of the input samples with floating point $p$ or $u$ are rounded to integers, as long as the same rounding operation is applied in both forward and inverse lifting steps. This implies that each lifting step can map integers to integers and is easily invertible.

Notice that each lifting step in the above factorization requires floating point multiplications. 
To avoid floating point multiplications, the lifting factors $p$ and $u$ can be approximated with rationals of the form $k/2^m$ ($k$ and $m$ are integers), which can be implemented with only addition and bitshift operations. Note that bitshift operation implicitly includes a rounding operation, which is necessary for mapping integers to integers, as discussed above. $k$ and $m$ can be chosen depending on the desired accuracy to approximate the plane rotation and the desired level of computational complexity. 

\subsection{I2i DCT}
\label{ssec:i2idct}
A significant amount of work on i2i transforms is done to develop i2i approximations of the discrete cosine transform (DCT). Although there are other methods, the most popular method, due its to lower computational complexity, is to utilize the factorization of the DCT into plane rotations and butterfly structures \cite{Chen,Loeffler,binDCT}. 

\begin{figure}[tbh]
  \begin{center}
    \includegraphics[trim=40 600 300 50,clip,width=0.8\linewidth]{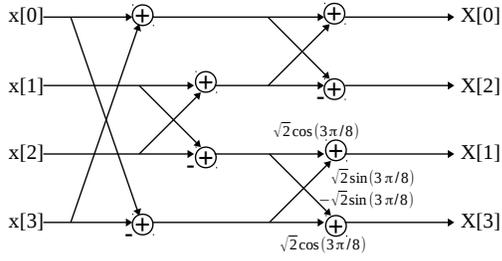}
    \caption{Factorization of 4-point DCT.}
    \label{fig:LoeffFact4}
  \end{center}
\end{figure}

Two well-known factorizations of the DCT into plane rotations and butterflies are the Chen's and Loeffler's factorizations \cite{Chen,Loeffler}. Loeffler's 4-point DCT factorization is shown in Figure \ref{fig:LoeffFact4}. It contains three butterflies and one plane rotation. Note that the output samples in Figure \ref{fig:LoeffFact4} need to be scaled by $1/2$ to obtain an orthogonal DCT.

The butterfly structures shown in Figure \ref{fig:LoeffFact4} map integers to integers because the output samples are the sum and difference of the inputs. They are also easily invertible by themselves and dividing the output samples by 2. 

\begin{figure}[thb]
  \begin{center}
    \includegraphics[trim=40 600 300 50,clip,width=0.8\linewidth]{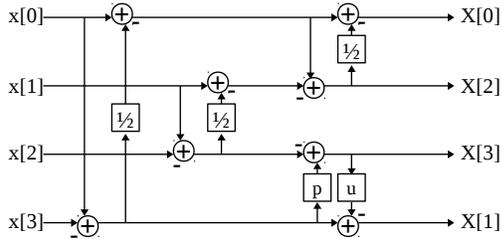}
    \caption{Lifting-based I2i DCT for lossless compression.}
    \label{fig:binDCT4}
  \end{center}
\end{figure}

The plane rotation in Figure \ref{fig:LoeffFact4} can be decomposed into three lifting steps or two lifting steps and two scaling factors as discussed in Section \ref{ssec:rotlif} to obtain integer-to-integer mapping. Using two lifting steps per plane rotation reduces the complexity. The two scaling factors can be combined with other scaling factors (if present) at the output, creating a scaled i2i DCT. The scaling factors at the output can be absorbed into the quantization stage in lossy coding. In lossless coding, all scaling factors can be omitted. However, care is needed when omitting scaling factors since for some output samples, the dynamic range may become too high when scaling factors are omitted. For example, in Figure \ref{fig:LoeffFact4}, the DC output sample becomes the sum off all input samples when scaling factors are omitted, however, it may be preferable that it is the average of all input samples. This can improve the entropy coding performance. Hence in lossless coding the butterflies of Figure \ref{fig:LoeffFact4} are replaced with lifting steps as shown in Figure \ref{fig:binDCT4} \cite{binDCT}.

\subsection{I2i DST}
\label{ssec:i2idct}
The odd type-3 DST has been shown to provide improved coding performance compared to the DCT for lossy coding of intra prediction residuals \cite{Chuo_j}. To the best of our knowledge, an i2i approximation of the DST has not appeared in the literature. To develop an i2i DST, we first approximate the DST with a cascade of plane rotations, and approximate these rotations with lifting steps to obtain an i2i DST.

\subsubsection{Approximation of DST with plane rotations}
\label{sssec:dstrot}
To the best of our knowledge an exact factorization of the DST into butterflies and/or plane rotations does not exist in the literature. We obtain an approximation of the DST by cascading multiple plane rotations using a modified version of the algorithm presented in \cite{Zeng_spl}.

Chen's algorithm \cite{Zeng_spl} can be briefly summarized as follows. A transform consisting of cascaded plane rotations is formed iteratively, where in each iteration, a pair of signal nodes is selected and the parameter ($\theta$) for the plane rotation is computed. The pair of nodes and the rotation parameter in each iteration are determined from the autocorrelation matrix of the input signal so that the coding gain of the transform after this plane rotation is maximized. In each iteration, the best rotation angle is determined from a closed form expression for each possible pair of nodes, and then the pair with the best coding gain is selected. The autocorrelation matrix is updated before moving to the next iteration. The iterations stop when a desired level of coding gain is achieved or a desired number of rotations are used.

\begin{figure}[thb]
  \begin{center}
    \includegraphics[trim=40 590 300 50,clip,width=0.8\linewidth]{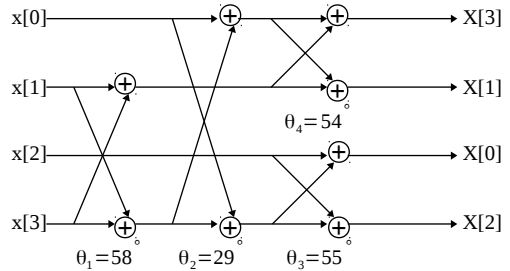}
    \caption{Cascaded plane rotations to approximate 4-point DST.}
    \label{fig:dstrot}
  \end{center}
\end{figure}

The auto-correlation of the input signal, which is the intra prediction residual signal ($x(i)$) in this paper, has been obtained in previous research based on modeling the image signal with a first-order Markov process and is given by $E[x(i)x(j)]=1+\rho^{|i-j|}-\rho^i-\rho^j$, where $\rho$ is the correlation coefficient of the Markov process \cite{Chuo_c,fatih_IBTbMP}. As $\rho \rightarrow 1$, the optimal transform for this auto-correlation is the odd type-3 DST \cite{Chuo_c,Chuo_j}. We use a $\rho$ value of $0.99$ in the auto-correlation expression and utilize a modified version of Chen's algorithm to obtain the cascade of plane rotations shown in Figure \ref{fig:dstrot} to approximate the 4-point DST with 4 rotations. 

\subsubsection{Scaled i2i DST approximation}
\label{sssec:dstlif}
\begin{figure}[tbh]
  \begin{center}
    \includegraphics[trim=40 600 300 50,clip,width=0.8\linewidth]{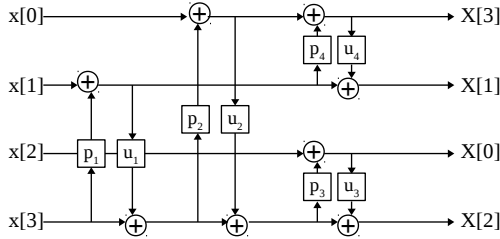}
    \caption{Lifting-based approximate I2i DST for lossless compression.}
    \label{fig:dstlif}
  \end{center}
\end{figure}

Each plane rotation in Figure \ref{fig:dstrot} can be decomposed into two lifting steps and two  scaling factors as discussed in Section \ref{ssec:rotlif}. Since we are interested in lossless coding, the two scaling factors are omitted and only lifting steps are used for representing each plane rotation, giving lifting implementation of a scaled DST shown in Figure \ref{fig:dstlif}. Note that omitting the scaling factors does not change the biorthogonal coding gain given in \cite{binDCT} but the rotation may become non-orthogonal. Finally, the lifting factors $p$ and $u$ are quantized to 3 bits for easy implementation of multiplications with addition and bitshift operations. Table \ref{tb:dstpu} lists the obtained lifting factors.

\begin{table}[tbh]
\centering
\caption{Lifting factors used in the i2i DST of Figure \ref{fig:dstlif}}
\label{tb:dstpu}
\begin{tabular}{cc|cc|cc|cc}
\hline 
 $p_1$ & $u_1$ & $p_2$ & $u_2$ & $p_3$ & $u_3$ & $p_4$ & $u_4$ \\ \hline
  -5/8 &   4/8 &  -3/8 &   2/8 &  -7/8 &   3/8 &  -5/8 &   4/8  \\ \hline
\end{tabular}
\end{table}

\subsection{I2i DCT and DST within HEVC}
\label{ssec:i2ihevc}
This paper uses the i2i DCT approximation shown in Figure \ref{fig:binDCT4} and the i2i DST approximation shown in Figure \ref{fig:dstlif} to explore i2i transforms in lossless coding within HEVC. The lifting factors $p$ and $u$ are quantized to 3 bits for easy implementation with addition and bitshift operations. The lifting factors in the i2i DCT (Figure \ref{fig:binDCT4}) are $p=3/8$ and $u=2/8$ \cite{binDCT}. The lifting factors in the i2i DST approximation (Figure \ref{fig:dstlif}) are given in Table \ref{tb:dstpu}.

The obtained i2i transforms are used along first the horizontal and then the vertical direction to obtain i2i 2D DCT and i2i 2D DST. These i2i 2D transforms are used in lossless compression to transform intra prediction residuals of luma and chroma pictures in only 4x4 transform units (TU). The transform coefficients are directly fed to the entropy coder without quantization. In larger TUs, the default HEVC processing is used. Notice that in lossless coding, the encoder choses 4x4 TUs much more frequently than other TUs. Exploring i2i transforms in larger TUs is left for future work.

\section{Experimental Results}
\label{sec:exp}

The i2i transforms are implemented into the HEVC version 2 Range Extensions (RExt) reference software (HM-15.0+RExt-8.1) \cite{HMref} to provide experimental results for the i2i transform approach. The following systems are derived from the reference software and compared in terms of lossless compression performance and complexity :
\begin{itemize}
  \itemsep0em
  \item HEVCv1
  \item HEVCv2
  \item i2iDCT
  \item i2iDCT+RDPCM
  \item i2iDST
  \item i2iDST+RDPCM.
\end{itemize}

The HEVCv1 system represents HEVC version 1, which just skips transform and quantization for lossless coding, as discussed in Section \ref{sec:intro}. The HEVCv2 system represents HEVC version 2, and includes all available RExt tools, such as RDPCM, reversing the scan order, a dedicated context model for the significance map and other tools \cite{HEVCv2,N0044} as discussed in Sections \ref{ssec:rdpcm} and \ref{ssec:entropy}.

The remaining two systems employ the i2i transforms discussed in Section \ref{ssec:i2ihevc}. The i2i transforms are used in intra coded blocks in only 4x4 transform units (TU). In larger TUs, the default processing of HEVC version 2 is used. 

In the i2iDCT and i2iDST systems, the RDPCM system of the reference software is disabled in 4x4 TUs, and is replaced with the 4x4 i2i 2D DCT and DST transform, respectively. In intra coding, these i2i transforms are applied to the residual TU of all intra prediction modes. 

In the i2iDCT+RDPCM and i2iDST+RDPCM systems, the i2i transform and RDPCM methods are combined in intra coding. In other words, in intra coding of 4x4 TUs, the RDPCM method of HEVC version 2 is used if the intra prediction mode is horizontal or vertical, and the i2i 2D DCT or DST transform is used for other intra prediction modes. 

For the experimental results, the common test conditions in \cite{commontest} are followed, except that only the first 32 frames are coded due to our limited computational resources. All results are shown in Table \ref{tb:res}, which includes average percentage ($\%$) bitrate reduction and encoding/decoding time of all systems with respect to HEVCv1 system for All-Intra-Main settings. 

\begin{table}[b]
\setlength{\tabcolsep}{3pt}
\centering
\caption{Average percentage ($\%$) bitrate reduction and encoding/decoding time of HEVCv2, i2iDCT, i2iDCT+RDPCM, i2iDST, i2iDST+RDPCM systems with respect to HEVCv1 in lossless coding for All-Intra-Main settings.}
\label{tb:res}
\begin{tabular}{l|ccccc}
\hline
      & HEVCv2  & i2iDCT & i2iDCT  & i2iDST  & i2iDST \\ 
      &         &        & +RDPCM  &         & +RDPCM \\ \hline \hline
Class A   & 7.2 &  9.9 & 10.9                &11.1     & 11.6      \\ \hline
Class B   & 4.7 &  4.3 &  5.0                & 5.4     & 5.8      \\ \hline
Class C   & 5.4 &  3.8 &  5.1                & 5.3     & 6.1      \\ \hline
Class D   & 7.6 &  6.4 &  8.1                & 7.6     & 8.7       \\ \hline
Class E   & 8.2 &  8.3 &  9.7                & 9.0     &10.2       \\ \hline
Average   & 6.4 &  6.3 &  7.5                & 7.5     & 8.3         \\ \hline
Enc. T. & $94\%$ & $96\%$ & $97\%$         & $96\%$ & $97\%$  \\ \hline
Dec. T. & $94\%$ & $94\%$ & $97\%$         & $94\%$ & $97\%$  \\ \hline
\end{tabular}
\end{table}

HEVCv2, i2iDCT, i2iDCT+RDPCM, i2iDST and $~$ i2iDST+RDPCM systems achieve $6.4$, $6.3$, $7.5$, $7.5$ and $8.3$ percent overall average bitrate reduction over HEVCv1, respectively. For Class A, which include sequences with the largest resolution (2560x1600), systems employing i2i transforms achieve significantly larger bitrate reductions than the HEVCv2 system. For the other classes, systems employing i2i transforms are typically slightly better than HEVCv2. 

Notice also that if the systems employing i2i transforms are compared, then systems employing the DST achieve on average around $1\%$ larger bitrate reduction than systems employing the DCT. This result is similar to the results comparing the DCT and DST in lossy intra coding \cite{HEVC}.



\section{Conclusions}
\label{sec:con}
This paper explored an alternative approach for lossless coding based on integer-to-integer (i2i) transforms within HEVC. I2i transforms map integers to integers without increasing the dynamic range at the output and were used in this paper to transform intra prediction residuals of luma and chroma pictures in only 4x4 transform units (TU). An i2i DCT from the literature and a novel i2i approximation of the DST were explored. Experimental results showed improved performance with respect to other major methods, such as RDPCM, in terms of both compression performance and complexity. 

%


\end{document}